\documentclass[a4paper,twocolumn,english,aps,prl,superscriptaddress]{revtex4}

\usepackage[T1]{fontenc}
\usepackage[latin1]{inputenc}
\usepackage{float}
\usepackage{graphicx,color}
\usepackage{bm}
\usepackage{bbm}
\usepackage{amsmath}
\usepackage{color}
\usepackage{amssymb}
\DeclareMathOperator{\sgn}{sgn}
\makeatletter

\usepackage{babel}
\makeatother
\begin{document}

\title{Position-momentum entangled photon pairs in non-linear wave-guides and transmission lines }

\author{Y. Sherkunov}
\address{National Graphene Institute, University of Manchester, Manchester, M13 9PL, UK}
\author{David M. Whittaker}
\address{Department of Physics and Astronomy, University of Sheffield, Sheffield S3 7RH, UK}
\author{Vladimir Fal'ko}
\address{National Graphene Institute, University of Manchester, Manchester, M13 9PL, UK}

\begin{abstract}
We analyse the correlation properties of light in non-linear wave-guides and transmission lines,  predict the position-momentum realization of EPR paradox for photon pairs in Kerr-type non-linear photonic circuits, and we show how  two-photon entangled states can be generated and detected.
\end{abstract}

\maketitle

Most modern communication systems are based on information transfer using light, and  quantum properties of light are already being used in securing information transfer protocols. This makes generation, controlled propagation and detection of entangled states of photons in optical circuits important elements in communication. Continuous-variable entanglement has been intensively studied  in view of developing such protocols \cite{Braunstein05,Reid09},  with vast majority of works focusing on quadrature components, where  entanglement has been observed between the amplitude and phase quadratures of squeezed light \cite{Ou92,Silberhorn01,Bowen03,Villar05}, continuous variable polarization entanglement   \cite{Bowen02,Oliver03,Ruifang07}, or transverse position-momentum entanglement in photon pairs produced by spontaneous parametric down-conversion process in crystals \cite{Howell04,Angelo04, Edgar12, Moreau14}. However, the implementation of continuous variable entanglement is mostly limited by free-space optical networks \cite{Masada15} requiring  increased complexity, high-precision alignment, and stability.

Here, we propose a theory describing photons entangled over continuous variables  in quantum circuits, whose elements are wave-guides or chains of high-quality resonators with strong Kerr-type non-linearity. In such systems the interaction between two photons leads to the four-wave mixing \cite{Wang01,Rarity05,Li105,Sharping06,Engin13,Carusotto99} resulting in the separation of bound pairs of photons, which propagation in the transmission line is position-correlated, from a continuous spectrum of two-photon states. The existence of bound photons discussed in this paper gives ways for a formation of strongly  position-momentum entangled photon states,  which  are collinear and  occupy a single transverse quantized wave-guide mode, making them a good candidate for  the implementation in quantum on-chip systems, in contrast with entangled pairs generated by conventional bulk-crystal entanglers.

The physical system where we expect the entangled photon states to appear include: (A) a Kerr-type non-linear single-mode wave-guide characterized by strong photon-photon coupling  \cite{Walker13, Walker15}, or (B) a chain of coupled non-linear resonators  \cite{Wallraff04, Schuster07, Hofheinz08, Devoret13, Yin13, Barends14}. For two photons with momenta $\hbar k_1=\hbar (k_0-\delta k)$ and $\hbar k_2=\hbar (k_0+ \delta k)$ and  dispersion 
\begin{eqnarray}
\omega_{k_0+\delta k} &\approx&  \omega_{k_0}+v \delta k +\beta \delta k^2/2,\label{omega}
\end{eqnarray}
where $v$ is the photon group velocity, the variation of the energy of a  photon pair 
\begin{eqnarray}
\Delta^{(2)}\omega=\omega_{k_0-\delta k} +\omega_{k_0+\delta k}-2\omega_{k_0} &\approx& \beta \delta k^2. \label{twophdisp}
\end{eqnarray}
As the photon-photon interaction conserves both energy and longitudinal momentum, the two-photon states propagating along the non-linear transmission line can be  described  by the Fock function
\begin{eqnarray}
|\psi\rangle_{2k_0}=\int dk_1 dk_2\delta (k_1+k_2-2k_0)f(k_1-k_2)|k_1,k_2\rangle. \label{Entwf}
\end{eqnarray}

\begin{figure}[ht]
\includegraphics[width=0.48\textwidth]{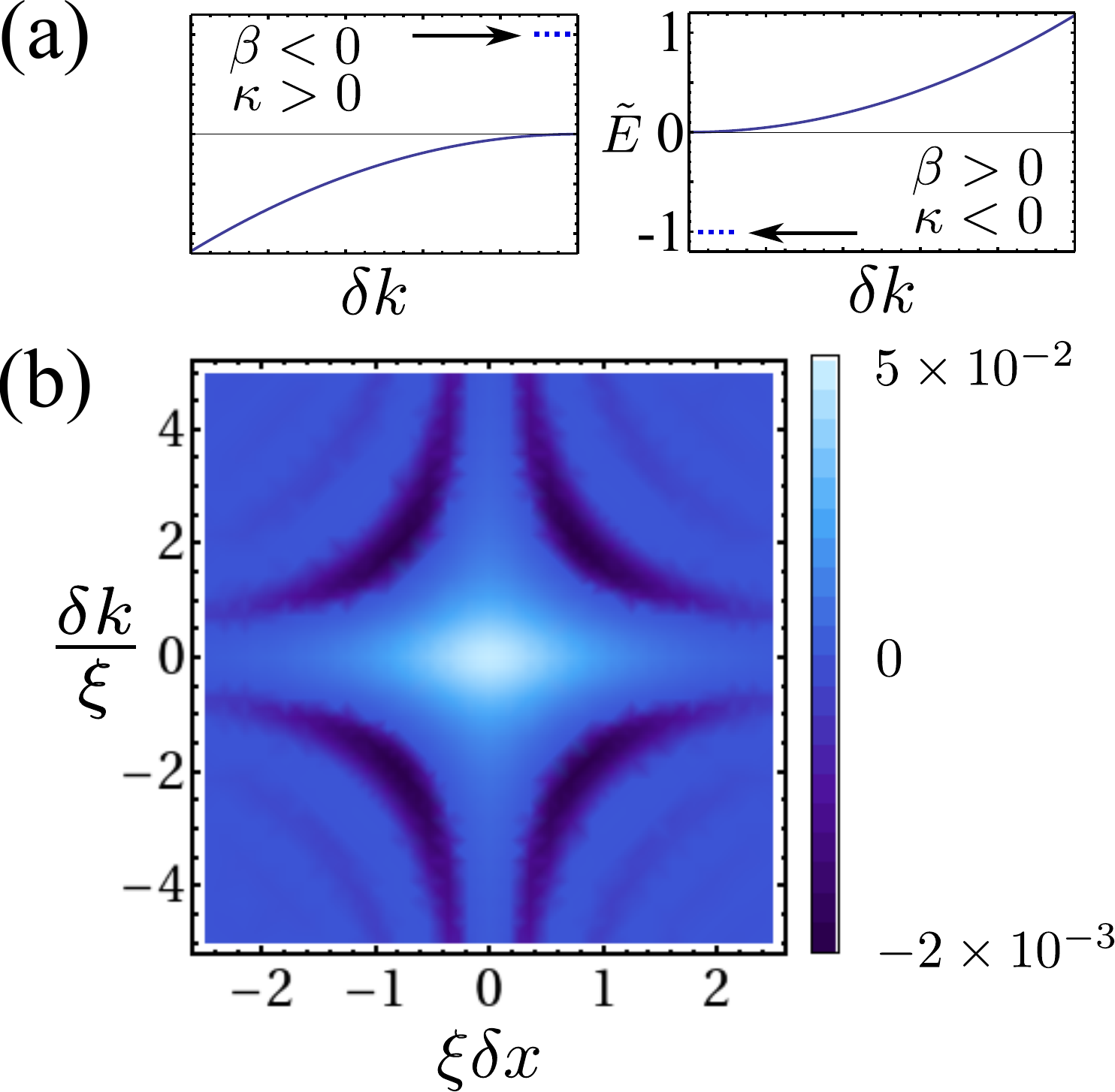}
\caption{Entangled two-photon states in non-linear wave-guides. (a) Spectrum of a two-photon state, $\tilde{E}=(E-2\omega_{k_0})|\beta|/\kappa^2$, with total momentum $2 k_0$  in a wave-guide with quadratic dispersion (\ref{omega}) for  $\beta<0$, $\kappa>0$ (left) and  $\beta>0$, $\kappa<0$ (right). Solid line corresponds to the continuous  spectrum, while the single eigenvalue corresponding to the entangled state is shown by  dashed line. (b) Wigner function of the two-photon entangled state. It takes negative values, which is a hallmark of non-Gaussian entangled states. }
\label{Fig1}
\end{figure}

(A)  To demonstrate the principle of position-momentum entanglement of photons in Kerr-nonlinear systems, we, first, consider the  entangled photon pairs in non-linear optical wave-guides. Classically, Kerr non-linearity in an isotopic medium manifests itself in the third-order polarisation 
\begin{eqnarray}
\mathbf{P}^{(3)(+)}=\chi^{(3)}[(\mathbf{E}^{(-)}\cdot \mathbf E^{(+)})\mathbf E^{(+)}+\alpha (\mathbf{E}^{(+)}\cdot \mathbf E^{(+)})\mathbf E^{(-)}],\nonumber
\end{eqnarray}
where $"+"$ and $"-"$ correspond to positive and negative frequency parts, $\mathbf E$ is electric field, $\chi^{(3)}$ is the susceptibility of the medium $\chi^{(3)}=\chi^{(3)}_{xyxy}$, $\alpha=\chi^{(3)}_{xxyy}/(2\chi^{(3)})$. Quantizing electromagnetic field, integrating over transverse degrees of freedom, and neglecting magneto-optical effects ($\alpha=0$) leading to entanglement over polarization degrees of freedom, we arrive at the following Hamiltonian ($\hbar=c=1$):

\begin{eqnarray}
H&=&H_0+H_{int},\;\;\; H_0=\sum_{k}\omega_k a_{k}^{\dagger}a_{k}\label{H} , \\
H_{int}&=&\frac{\kappa}{L} \sum_{k_1,k_2,k_3,k_4}\delta(k_1+k_2,k_3+k_4) a_{k_4}^{\dagger}a_{k_3}^{\dagger} a_{k_1}a_{k_2},\nonumber 
\end{eqnarray}
where $a_k$ ($a_k^{\dagger}$)  is the annihilation (creation) operator of a photon with longitudinal momentum $k$ and energy $\omega _k$, $L$ is the length of the system. The non-linear term $H_{int}$ in Eq. (\ref{H}) describes photon-photon interaction with coupling  $\kappa=\frac{\pi\omega^2 \chi ^{(3)}}{2n_r^4A\epsilon_0}$, where $n_r$ is  refraction index,  $A$ is the area occupied by the wave-guide mode and $\epsilon_0$ is the vacuum permittivity.

Hamiltonian (\ref{H}) can be diagonalized exactly in the case of $\Delta^{(2)}\omega\propto\delta k^2$ \cite{Thacker81}. We consider a sector of the Hilbert space, which consists of all the two-photon states with the total pair momentum  $2 k_0$ and assume the  effective mass approximation for the wave-guide dispersion given by Eq. (\ref{omega}). In the coordinate domain, $a_x=\frac{1}{\sqrt{L}}\sum_k e^{i(k-k_0)x}a_k$, the Hamilton Eq. (\ref{H}) takes the form: 
\begin{equation*}
H=\int dx \left(\omega_{k_0}a_x^{\dagger}a_x-iv a_x^{\dagger}\partial_xa_x-\frac{1}{2}\beta a_x^{\dagger}\partial^2_xa_x\right)
\end{equation*}
\begin{equation}
+\frac{1}{2}\int dx_1dx_2 a_{x_1}^{\dagger}a_{x_2}^{\dagger}U(x_1-x_2)a_{x_2}a_{x_1},\label{Hsp}
\end{equation}  
where $U(x_1-x_2)=2\kappa\delta(x_1-x_2)$.  For a two-photon state, described by the wave-function 
\begin{equation*}
|\psi\rangle=\int dx_1dx_2 f(x_1,x_2) a_{x_1}^{\dagger}a_{x_2}^{\dagger}|0\rangle,
\end{equation*}
this leads to the following Schr\"odinger equation:  
\begin{widetext}
\begin{eqnarray}
[2\omega_{k_0}-iv(\partial_{x_1}+\partial_{x_2})-\frac{1}{2}\beta( \partial_{x_1}^2+ \partial_{x_2}^2)
+2\kappa\delta(x_1-x_2)]f(x_1,x_2)=Ef(x_1,x_2),\label{schr}
\end{eqnarray}
\end{widetext}
where $E$ is the energy of a two-photon state. Equation (\ref{schr}) has scattering state solutions, which  correspond to the continuous spectrum of non-interacting photons with energies given by Eq. (\ref{twophdisp}) (See Fig. \ref{Fig1}(a)). When  the curvature of the wave-guide dispersion $\beta$ and the photon-photon coupling constant $\kappa$ are of opposite signs, $\beta\kappa<0$, there exists a bound state solution with
\begin{eqnarray}
f(x_1,x_2)=\sqrt{\frac{\xi}{2L}}\exp\left [-|x_1-x_2|\xi\right], \; \xi=|\kappa/\beta|\label{contins}
\end{eqnarray}
The energy of this state is split from the continuum of weakly correlated scattering states, as we show in Fig. \ref{Fig1}(a), and it is given by
\begin{eqnarray}
E_b=2\omega_{k_0}-\kappa^2/\beta ,\label{Esq}
\end{eqnarray}  
as expected from binding of a one-dimensional massive particle to an attractive $\delta$-functional potential well \cite{landau1977quantum}. In the momentum domain, the two-photon bound state  wave-function is given by Eq. (\ref{Entwf})
with  
\begin{eqnarray}
f(k_1-k_2) &=& \frac{8\xi^{3/2}}{\sqrt{2L}[(k_1-k_2)^2+4\xi^2]}.\label{kent}
\end{eqnarray}

The state (\ref{kent}) can be characterised  by the  Wigner function defined as the expectation value  $W(x_1,k_1;x_2,k_2)=\pi^{-2}\langle \psi| \Pi(x_1,k_1)\otimes \Pi(x_2,k_2)|\psi\rangle$ of the parity operator $\Pi(x,k)=\int d\zeta e^{-2ix\zeta}a_{k+\zeta}^{\dagger}|0\rangle\langle 0|a_{k-\zeta}$. After straightforward calculations, one can find
\begin{widetext}
\begin{eqnarray}
 W(x_1,k_1;x_2,k_2)=\frac{\xi^2e^{-2\xi|\delta x|}}{2\pi^2(\delta k^2+\xi^2)}\left(\cos(2\delta k|\delta x|)+\frac{\xi}{\delta k}\sin (2\delta k|\delta x|)\right)\delta (k_1+k_2;2k_0),
\end{eqnarray} 
\end{widetext}
where $\delta x=x_1-x_2$. This function is negative for $\cos(2\delta k|\delta x|)+\xi/\delta k\sin(2\delta k|\delta x|)<0$, as shown in Fig. \ref{Fig1}(b), which implies that the state (\ref{kent}) is entangled in position-momentum degrees of freedom \cite{Douce13}. Moreover, for $\xi\rightarrow\infty$, the two-photon wave-function approaches the ideal Einstein-Podolsky-Rosen state (see Appendix A). 

Alternatively, to demonstrate that the state (\ref{kent}) is entangled in position-momentum degrees of freedom, one can find  the uncertainties $\Delta(x_1-x_2)$ and $\Delta(k_1+k_2)$ calculated over the joint probability distributions $P(x_1,x_2)$ and $P(k_1,k_2)$ respectively, for which, the separability criterion \cite{Mancini02, Reid09, Duan00}:
\begin{eqnarray}
[\Delta (x_2-x_1)]^2[\Delta (k_2+k_1)]^2\geq 1, \label{EP}
\end{eqnarray}
can be applied. Although, the states for which the inequality (\ref{EP}) is violated are inseparable, they do not necessarily lead to EPR paradox. In order for an EPR paradox to arise, correlations must violate a more strict inequality  \cite{Reid89}:
\begin{eqnarray}
[\Delta (x_2-x_1)]^2[\Delta (k_2+k_1)]^2\geq 1/4, \label{EP1}
\end{eqnarray}
which  can be accessible   experimentally \cite{Moreau14}.
Assuming that the system is driven by a Gaussian beam of width $W_p$ in momentum space, we find that the entangled photon states are described by the wave-function (\ref{Entwf}) with the $\delta$-function substituted by the Gaussian $\delta (k_1+k_2-2k_0)\rightarrow (2/\pi)^{1/4}(L/2\pi W_p)^{1/2}\exp[-(k_1+k_2-2k_0)^2/W_p^2]$ and $f(k_1-k_2)$ given by Eq. (\ref{kent}). For the case of narrow Gaussian beam with $W_p\ll \xi$, we find $[\Delta (x_2-x_1)]^2[\Delta (k_2+k_1)]^2=\frac{1}{8}\left(\frac{W_p}{\xi}\right)^2$, which violates both inequalities (\ref{EP}) and  (\ref{EP1}).

\begin{figure}[h] 
\includegraphics[width=0.42\textwidth]{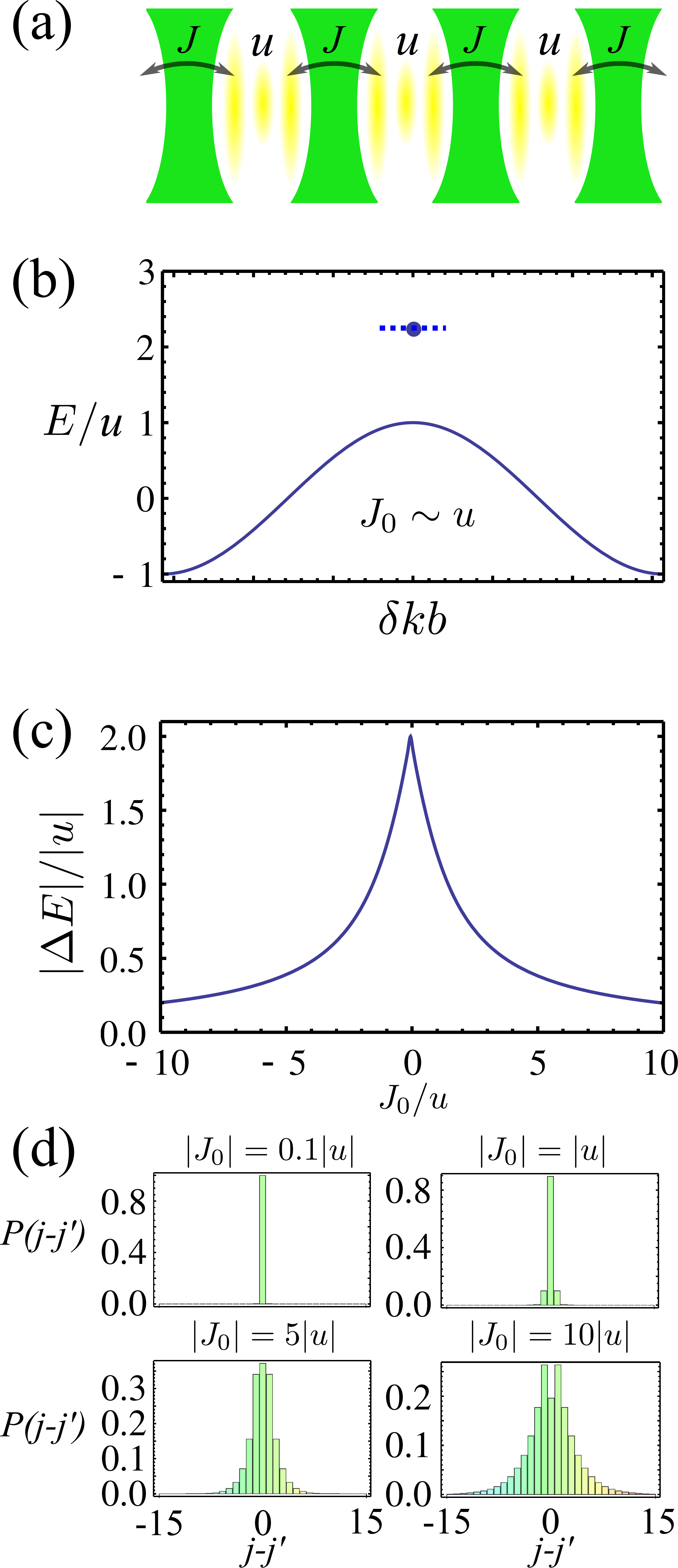}
\caption{ Entangled photon pairs in chains of coupled non-linear resonators. (a) Schematic of the set-up; (b) An example of two-photon spectrum for $J_0\sim u$ and $\omega_c=0$; (c) Split of the bound-state eigenvalue from the spectrum edge, $\Delta E$, as a function of $J_0=4J\cos(k_0a)$; (d) Joint probability distribution $P$ for various ratios $J_0/u$.  }
\label{Fig2}
\end{figure}

(B) Another system, where entangled photon pairs may appear is a chain (with period $b$) of $N$ coupled resonators illustrated in Fig. \ref{Fig2}(a). Here, each optical circuit element is characterized by a single photonic mode of frequency $\omega_c$ and non-linear on-site photon-photon interaction $u$. The photons can hop between the neighbouring cavities with an amplitude $J$, which can be described  by the Bose-Hubbard model: 
\begin{eqnarray}
H=\sum_j(\omega_c a_j^{\dagger}a_j+J(a_{j+1}^{\dagger}a_j+h.c.)+ua_{j}^{\dagger}a_j^{\dagger}a_ja_{j}),\label{BH}
\end{eqnarray}
where $a_j$ ($a_j^{\dagger}$) is the annihilation (creation) operator of a photon on site $j$. Hamiltonian (\ref{BH}) can also be diagonalized exactly in the two-particle subspace of the Hilbert space \cite{Valiente08,lewenstein2012ultracold}. Using the periodic boundary conditions for closed chain ($j=N+1\equiv 1$),  the system is described by the following single photon dispersion, $\omega_k$
\begin{eqnarray}
\omega_k&=&\omega_c+2J\cos(kb), \nonumber\\
k&=&2\pi n/(Nb),\; n=0,1,2,...,N-1.\label{dispcav}
\end{eqnarray}  
For the two-photon states, 
\begin{eqnarray}
|\psi\rangle = \sum_{j'\geq j}f(j'-j)e^{ik_0b(j+j')}a^{\dagger}_ja^{\dagger}_{j'}|0\rangle,\nonumber
\end{eqnarray}
the Schr\"{o}dinger equation, $H|\psi\rangle =E |\psi\rangle$, is equivalent to
\begin{flalign}
&J_0f(1)=2(E-2\omega_c-2u)f(0), &\label{shr}\\
&J_0f(j+1)=2(E-2\omega_c)f(j)-(1+\delta_ {j,1})J_0f(j-1).& \nonumber
\end{flalign}
Here, $J_0=4J\cos(k_0b)$ is the energy of two non-interacting photons each with quasi-momentum $ k_0$. 

The scattering-state  solution  has energy of non-interacting photon pair $E_{sc}=2\omega_c+J_0\cos(\delta kb)$  and wave-function 
\begin{eqnarray}
f_{sc}(j)=2\left(\cos(\delta kjb)-\frac{2u\sin(\delta kjb)}{J_0\sin (\delta kb)}\right )f(0).\nonumber \label{scatt}
\end{eqnarray}
Moreover, Eq. (\ref{shr}) has a bound-state solution independently of the signs of the coupling constant $u$ and curvature of the spectrum. This state has energy 
\begin{eqnarray}
E_b=2\omega_c+\sgn (u) \sqrt{J_0^2+4u^2} \label{Earr}
\end{eqnarray}
and wave-function 
\begin{eqnarray}
f_{b}(j) &=& 2\sqrt{\frac{(1-\eta^2)}{N(1+3\eta ^2)}}\left(\eta^{|j|}-\frac{\delta_{j,0}}{2}\right),\nonumber\\
\eta &=& \frac{1}{J_0}\left(-2u+\sgn(u)\sqrt{J_0^2+4u^2}\right), \label{fjj}
\end{eqnarray}
In the wave-number representation, this reads
\begin{eqnarray}
|\psi\rangle=\sqrt{\frac{(1-\eta^2)^{3}}{N(1+3\eta^2)}}\sum_{k_1 k_2}\frac{\delta(k_1+k_2;2k_0)a_{k_1}^{\dagger}a_{k_2}^{\dagger}}{1-2\eta\cos\left[\frac{(k_1-k_2)b}{2}\right]+\eta^2}|0\rangle. \nonumber
\end{eqnarray}
This state is separated from the quasi-continuum of scattering states as we show  in Fig. \ref{Fig2}(b).

For strong photon-photon coupling, $|u|\gg |J_0|$, Eq. (\ref{fjj}) yields 
\begin{eqnarray}
f_b(j)&=& \frac{2}{\sqrt{N}}\left(1-\frac{\delta_{j,0}}{2}\right)\left(\frac{J_0}{4u}\right)^{|j|},\label{farrayu}\\
E_b&=&2\omega_c+2u+J_0^2/(4u). \nonumber
\end{eqnarray}
In this case, the photon-photon correlation length is small, so the two correlated photons tend to occupy the same resonator, with their energy approaching the on-site interaction energy $2u$ independent of weather the interaction is repulsive or attractive (see Figs. \ref{Fig2}(c) and (d)). It is worth mentioning that, for $\eta\rightarrow 0$, the wave-function (\ref{farrayu}) mimics a perfect EPR pairs (see Appendix A).

In the case of weak photon-photon coupling,  $|J_0|\gg |u|$, the correlated photon pair has large correlation radius  and small energy separation from continuum of scattering states,  Figs. \ref{Fig2}(c) and (d). In this case  we find  
\begin{widetext}
\begin{eqnarray}
f_b(j)= \frac{2}{\sqrt{N}}\left(1-\frac{\delta_{j,0}}{2}\right)\left[\sgn(uJ_0)\left(1-\frac{|2u|}{|J_0|}+\frac{2u^2}{J_0^2}\right)\right]^{|j|}\label{farrayJ}
\end{eqnarray}
\end{widetext}
and 
\begin{eqnarray}
E_b=2\omega_c+\sgn(u)(|J_0|+2u^2/|J_0|^2).\label{EarrayJ}
\end{eqnarray}
in agreement with the continuous model Eq. (\ref{contins}) \cite{footnote2}.  The transition between the two extremes of strongly and weakly interacting photons is shown in Fig. \ref{Fig2}(d). 

Experimentally,  entangled states discussed above could  be generated by applying a coherent pump, such as monochromatic laser beam, to the chain of resonators. The results of numerical simulations of the generation of entangled photon states in a closed chain of three lossy cavities driven by a weak coherent laser source shown in Fig. \ref{Fig3} (see Appendix B) suggest that the most effective generation of two-photon entangled states occurs when the pump frequency, $\omega_p$, satisfies  the resonant condition for  bound photon pairs, $\omega_p=E_b/2$, for which the zero-time delay on-site correlation function, $g^{(2)}_{jj}=\langle (a_j^{\dagger})^2a_j^2\rangle/\langle a_j^{\dagger}a_j\rangle^2$, takes its maximum value \cite{Sherkunov14,Sherkunov16}. Single-photon resonance takes place at $\omega_p=\omega_c+J_0/2$, which can be seen as a maximum of on-site number of photons, $N_j$, and minimum of $g^{(2)}_{jj}$. This corresponds to the generation of scattering states. The momentum of two-photon state, $2k_0$, is determined by site-dependent phase of  pumping.  

\begin{figure}[h] 
\includegraphics[width=0.45\textwidth]{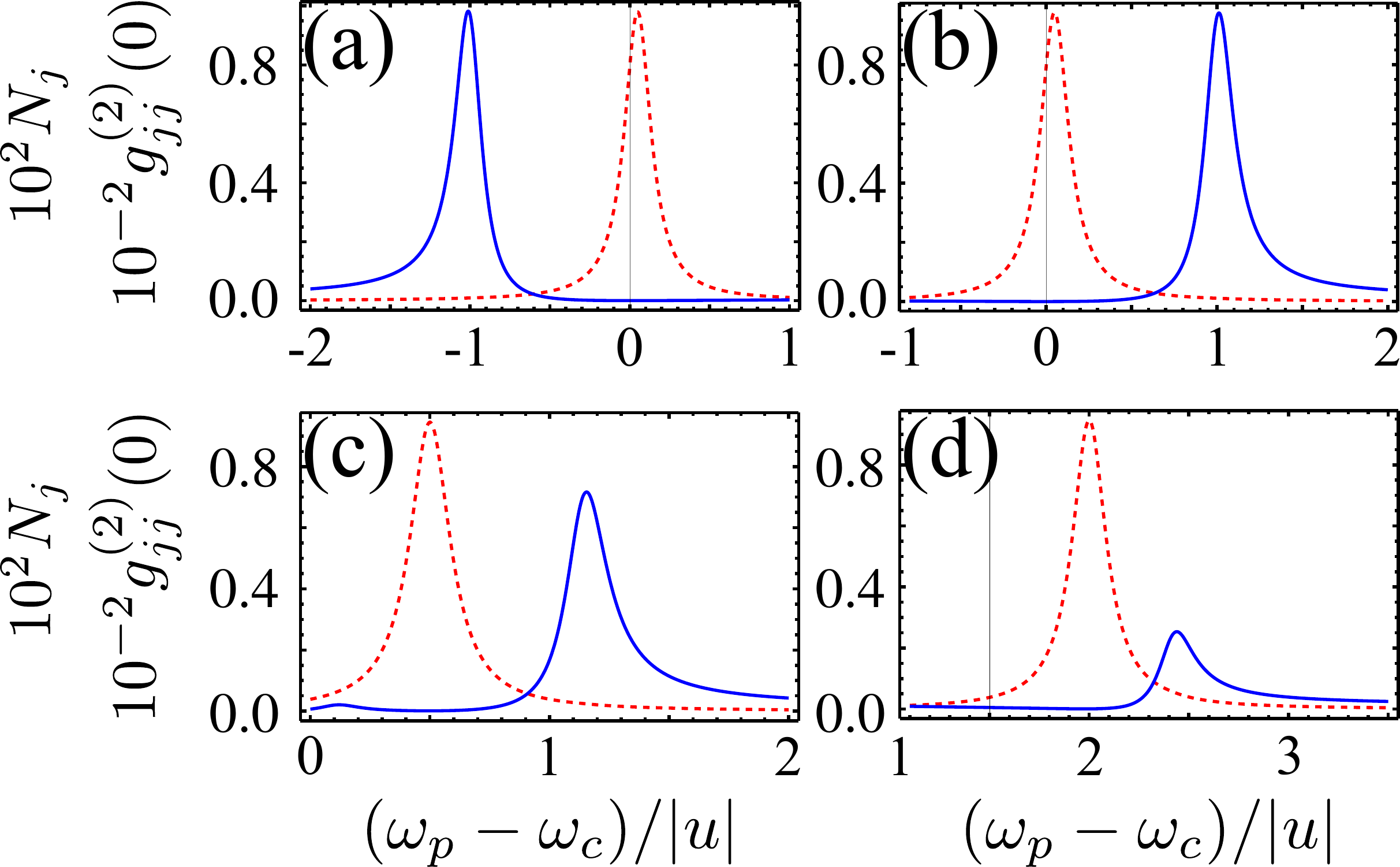}
\caption{Steady-state average number of photons $N_j$ (dashed line) and on-site zero-time delay correlation function $g_{jj}^{(2)}(0)$ (solid line) in resonator $j$, as a function of pumping frequency, $\omega_p$, for  coherently pumped short chain (three resonators) with pumping amplitude $F=10^{-2}|u|$, photon decay rate $\gamma=0.1|u|$, and pair momentum $2k_0=0$. (a) $J_0=0.1|u|$, $u<0$; (b) $J_0=0.1|u|$, $u>0$; (c) $J_0=|u|$, $u>0$; (d) $J_0=4|u|$, $u>0$. Note that far from the resonances $g^{(2)}_{jj}(0)\rightarrow 1$.   }
\label{Fig3}
\end{figure} 

To conclude, we have shown that photon pairs entangled over continuous variables such as position and momentum can be generated in quantum systems whose elements are either non-linear wave-guides or chains of optical or microwave non-linear resonators due to photon-photon interaction stemmed by Kerr-type non-linearity. In the case of strong photon-photon interaction, the generated states give good approximation to the EPR state. 

The theory is formulated independently of the frequency range of photons used. It can be applied to  visible-range polaritonic wave-guide systems, where the non-linearity is due to exciton-exciton interaction \cite{Timofeev14}. It is also applicable to   microwave-frequency superconducting transmission lines of high-quality resonators coupled to qubits \cite{Houck12}.  The latter system is more appealing because of controllability of its parameters, high Q-factor (low losses) and a stronger non-linearity  \cite{Houck12},  which enables one to reach the regime with $u/\gamma\gg1$ making the effect of losses  on the EPR correlations negligible. Strong non-linearity and low losses  can also be  achieved in the systems, where  atoms in electromagnetically induced transparency regime  \cite{Imamo97, Hartmann07} are coupled to     microcavities with high quality factors, $Q$, such as toroidal ($Q>10^8$) \cite{Armani03} or microrod ($Q>10^9$)  \cite{Del13} resonators. In these systems, the strength of non-linearity can reach $u\sim 1.25 \times 10^7 s^{-1}$, while the losses can be as low as $\gamma\sim 2 \times 10^{5}s^{-1}$ \cite{Armani03,Aoki06,Spillane05,Hartmann07}, hence reaching the desirable regime.

The position-momentum entangled pairs discussed in this paper, in comparison with the ones generated by  conventional bulk-crystal entanglers, are collinear and predominantly occupy a single transverse quantized wave-guide mode, which offers a potential for the implementation in quantum on-chip circuits.  Experimentally, the entangled states could be generated by applying coherent pumping to the system with frequency satisfying the resonance condition for bound photon pairs. These states could be accessed by measuring the two-photon Wigner function  in a Hong-Ou-Mandel type experiment \cite{Douce13}. It can play a role of an entanglement witness taking negative values, as shown in Fig. \ref{Fig1}(a),  for non-Gaussian entangled states.  We have  also demonstrated that EPR correlations of the states discussed in this paper would lead to violation of experimentally accessible \cite{Moreau14} criterion (\ref{EP1}).

We thank D. Krizhanovskii, E. Cancellieri and M. Skolnick for useful discussions. This
work was supported by EPSRC Programme Grant
EP/J007544.

\appendix
\section{Appendix A}
 \renewcommand{\theequation}{A\arabic{equation}}
  \setcounter{equation}{0}  
In the  case of the wave-guide with linear dispersion ($\beta =0$), one can find $f(k_1-k_2)=const$. This is  the ideal position-momentum entangled state proposed by Einstein, Podolsky and Rosen (EPR) {\cite{EPR}}, in which position, $x$, and momenta are perfectly (anti-) correlated: 
{\begin{eqnarray}
|\psi \rangle=\int d(\delta k)|k_0+\delta k,k_0-\delta k\rangle=\int dx e^{2ik_0x}|x,x\rangle. \label{EPR}
\end{eqnarray}}
To demonstarte this, we rewrite the Hamiltonian (\ref{H})  as an {$(N+1)/2\times (N+1)/2$}   matrix  {$H =2\omega_{k_0}\mathbb{I}+
\frac{2\kappa}{L}\left( \begin{array}{cccc}
1 & \sqrt{2} & \sqrt{2} & ...  \\
\sqrt{2} & 2& 2 & ... \\
 \sqrt{2} & 2 & 2 & ...\\
... & ... & ... & ... \end{array} \right)$} in the basis spanning {$(N+1)/2$} two-photon states with total momentum {$2 k_0$}. It can be diagonalised exactly yielding the following eigenvalues: bound-state eigenvalue  {$E_{b}=2\omega_{k_0}+\kappa k_{max}/\pi$}, where {$k_{max}$} is the maximum wave-number corresponding to the break-down of the linear approximation,  and continuous spectrum eigenvalues  {$E_{\delta k}=2\omega_{k_0}$}. The wave functions corresponding to the bound-state wave-function is found to be {$\psi_\propto \frac{1}{\sqrt{2}} (a^{\dagger}_{k_0})^2|0\rangle+\sum_{\delta k>0}a^{\dagger}_{k_0+\delta k}a^{\dagger}_{k_0-\delta k}|0\rangle $} and the continuous spectrum wave-functions are {$\psi_{\delta k}\propto -\sqrt{2} (a^{\dagger}_{k_0})^2|0\rangle+a^{\dagger}_{k_0+\delta k}a^{\dagger}_{k_0-\delta k}|0\rangle$}. In coordinate domain, the bound-state two-photon wave-function is {$\psi_\propto \int dx_1dx_2e^{ik_0(x_1+x_2)}\delta(x_1-x_2)a^{\dagger}_{x_1}a^{\dagger}_{x_2}|0\rangle$. 

\section{Appendix B}
\renewcommand{\theequation}{B\arabic{equation}}
  \setcounter{equation}{0}  
The density matrix, {$\rho$}, describing the evolution of photons in  three coherently driven lossy cavities obeys the master equation :
{\begin{eqnarray}
\partial_t\rho&=&-i[H+H_{p},\rho]\nonumber\\
&+&\gamma \sum_{j=1,2,3}(2 a_j\rho a_j^{\dagger}- a_j^{\dagger}  a_j\rho -\rho  a_j^{\dagger} a_j),\label{mastereq}
\end{eqnarray}} 
where {$\gamma$} is the photon decay rate, {$H_p=\sum_j (F_j(t)a_j^{\dagger}+h.c.)$} and {$F_j(t)=Fe^{-i\omega_pt+i\psi_j}$} describes coherent pumping with amplitude {$F$}, frequency {$\omega_p$} and phase {$\psi_j$}. The latter determines the momentum of generated photons. By finding the steady-state solution of the master equation (\ref{mastereq}) for the density matrix determined in the Fock space of photon states with different occupation numbers of the three cavities, {$\rho=\sum P(m_1,m_2,m_3,n_1,n_2,n_3)|m_1,m_2,m_3\rangle \langle n_1,n_2,n_3|$}, we evaluate the numbers {$N_j$}, of photons in each cavity as well as zero time-delay on-site pair correlation function {$g^{(2)}_{jj}(0)$}. We assumed {$\psi_j=0$}.

\bibliography{ybs.bib}{}

\begin{thebibliography}{48}
\bibitem[{foo()}]{footnote2}
\bibinfo{note}{Indeed, taking $N\rightarrow\infty$ and expanding Eq.
  (\ref{dispcav}) in the vicinity of $k\approx k_0$ up to the second order in
  $\delta k$, one arrives at Eq. (\ref{omega}) with $\omega_{k_0}=J_0/2$,
  $v=2Jb\sin(k_0b)$, and $\beta=-J_0b^2/2$. Substituting this results into Eqs.
  (\ref{contins}) and (\ref{Esq}), taking into account that for
  $N\rightarrow\infty$ the coupling constants of both models are related as
  $\kappa=ub$, and expanding upto the second order in $u/J_0$, we arrive at
  Eqs. (\ref{farrayJ}) and (\ref{EarrayJ}).}
\end{thebibliography}
\bibliographystyle{apsrev}
\end{document}